\documentclass[12pt]{article}
\DeclareMathAlphabet{\scr}{U}{rsfs}{m}{n}
\usepackage{latexsym,url}
\usepackage[mathscr]{eucal}
\usepackage{amsfonts}
\usepackage{amscd}
\usepackage{cite}         
\usepackage{amssymb}
\usepackage{color}
\usepackage{colordvi}
\usepackage[centertags]{amsmath}
\usepackage{enumerate}
\usepackage{graphicx}
\usepackage{booktabs}
\usepackage{hyperref}

\newcommand{\cleqn}{\setcounter{equation}{0}}
\newcommand{\newc}{\newcommand}
\newc{\be}{\begin{equation}}
\newc{\ee}{\end{equation}}
\newc{\bea}{\begin{eqnarray}}
\newc{\eea}{\end{eqnarray}}
\newc{\ben}{\begin{equation*}}
\newc{\een}{\end{equation*}}
\newc{\bean}{\begin{eqnarray*}}
\newc{\eean}{\end{eqnarray*}}
\newc{\ol}{\overline}
\newc{\wt}{\widetilde}
\newc{\bs}{\boldsymbol}
\newc{\m}{\mathcal}
\newc{\la}{\lambda}
\newc{\lra}{\longrightarrow}
\newc{\vp}{\varphi}
\newc{\ti}{\tilde}
\newc{\VEV}[1]{\langle#1 \rangle}

\begin{document}

\title{\hfill ~\\[-30mm]
          \hfill\mbox{\small  QFET-2017-05}\\[-3.5mm]
          \hfill\mbox{\small  SI-HEP-2017-06}\\[13mm]
       \textbf{Minima of
multi-Higgs potentials with triplets of $\Delta(3n^2)$ and $\Delta(6n^2)$
}}

\author{
Ivo de Medeiros Varzielas$\,^{a,b}\,$\footnote{E-mail: {\tt ivo.de@udo.edu}}
,~~
Stephen F. King$\,^a\,$\footnote{E-mail: {\tt king@soton.ac.uk}}
,\\[2mm]
Christoph Luhn$\,^{c}\,$\footnote{E-mail: {\tt christoph.luhn@uni-siegen.de}}
,~~
Thomas Neder$\,^{a,d}\,$\footnote{E-mail: {\tt neder@ific.uv.es}}\\[8mm]
$^a$\,\it{\small School of Physics and Astronomy, University of Southampton,}\\
\it{\small SO17 1BJ Southampton, United Kingdom}\\
$^b$\,\it{\small CFTP, Departamento de F\'{\i}sica, Instituto Superior T\'{e}cnico,}\\
\it{\small Universidade de Lisboa,
Avenida Rovisco Pais 1, 1049 Lisboa, Portugal}\\
$^c$\,\it{\small Theoretische Physik 1, Naturwissenschaftlich-Technische
    Fakult\"at, Universit\"at Siegen,}\\
\it{\small Walter-Flex-Stra{\ss}e 3, 57068 Siegen, Germany}\\
$^d$\,\it{\small AHEP  Group,  Institut  de  F\'{\i}sica  Corpuscular (IFIC) --- C.S.I.C./Universitat  de  Val\`{e}ncia,}\\
\it{\small Parc  Cient\'{\i}fic  de  Paterna, C/  Catedr\'{a}tico  Jos\'{e}  Beltr\'{a}n,  2  E-46980  Paterna  (Valencia), Spain}\\
}

\date{\today}

\maketitle

\begin{abstract}
\noindent  
{
We analyse the minima of scalar potentials for multi-Higgs models 
where the scalars are arranged as either one triplet or two triplets of the discrete symmetries
$A_4$, $S_4$, $\Delta(27)$, $\Delta(54)$, as well as $\Delta(3n^2)$ and $\Delta(6n^2)$ with $n>3$.
The results should be useful for both multi-Higgs models involving 
electroweak doublets and multi-flavon models involving electroweak singlets,
where in both cases the fields transform as triplets under some non-Abelian discrete symmetry.}
\end{abstract}
\thispagestyle{empty}
\vfill





\section{Introduction}

Following the discovery of the Higgs boson of the Standard Model (SM), it remains an intriguing possibility
that there are more scalar bosons waiting to be discovered.
Indeed, many extensions beyond the Standard Model include additional scalars, whether electroweak $SU(2)_L$ doublets in multi-Higgs doublet models or $SU(2)_L$ singlets typically found for example in flavour models in order to break some family symmetry. Given this, it is important to catalogue the minima of potentials including several scalars. In general this 
is a technically difficult task, which simplifies somewhat for simple cases where the potential
is controlled by a large discrete symmetry.

In this paper we consider potentials of scalars which transform as triplets under various non-Abelian discrete family symmetries.
The potentials we consider therefore involve 
of up to six scalar $SU(2)_L$ doublets or singlets.
We follow a progressive method that relies on considering which degrees of freedom become physical as the symmetry of the potential decreases when adding terms, we find a list of minima (not necessarily exhaustive) for potentials with one and two scalar triplets of $A_4$, $S_4$, $\Delta(27)$, $\Delta(54)$, and $\Delta(3n^2)$ and $\Delta(6n^2)$ with $n>3$. 
These symmetries \cite{Luhn:2007uq, Escobar:2008vc} are typically used in multi-Higgs doublet models \cite{Branco:1983tn, deMedeirosVarzielas:2011zw, Fallbacher:2015rea, Varzielas:2015joa, Pascoli:2016eld} and as family symmetries \cite{King:2013vna, Ding:2014ora}. The explicit CP properties of all these potentials were analysed recently with invariant methods \cite{Varzielas:2016zjc}.
We start with potentials of one triplet, many of which had been studied in \cite{Ivanov:2012ry, Ivanov:2012fp}, and their minima found in \cite{Ivanov:2014doa} with a geometric method developed in \cite{Ivanov:2010wz, Degee:2012sk}. Minimisation methods are also reviewed in \cite{Ivanov:2017dad}. We then consider the two triplet cases based on the results of the one triplet cases. Minima related by symmetries of the potential form sets of related Vacuum Expectation Values (VEVs) referred to as orbits. When the symmetry of a potential is decreased by adding terms, this has the effect of splitting larger orbits into several smaller orbits. In addition, we mostly disregard the magnitude of the VEVs and focus mainly on their alignments.

The layout of the remainder of the paper is as follows.
We start by going through the potentials with one triplet and list their minima throughout Section \ref{sec:one}, with $\Delta(6n^2)$ and $\Delta(3n^2)$ with $n>3$ on Section \ref{sec:6n3n1}, $S_4$ in \ref{1xS4_vev_section}, $A_4$ in \ref{1xA4_vev_section}, $\Delta(54)$ and $\Delta(27)$ in \ref{sec:54271}. We use these results to then find minima for two triplet potentials in Section \ref{sec:two}, with $\Delta(6n^2)$ and $\Delta(3n^2)$ with $n>3$ respectively in Sections \ref{sec:6n22} and \ref{sec:3n22}, $S_4$ and $A_4$ in \ref{sec:S4PPVEVs} and \ref{sec:A4PPVEVs}, $\Delta(54)$ and $\Delta(27)$ in \ref{sec:D54PPVEVs} and \ref{sec:D27PPVEVs}. We conclude in Section \ref{sec:con}.

\cleqn

\section{Potentials and VEVs with one triplet \label{sec:one}}

We use cycl.\ to denote the cyclic
permutations, and h.c.\ to indicate the hermitian conjugate. In addition to the discrete symmetries, the potentials of $SU(2)_L$ singlets are invariant under additional $U(1)$ symmetries to eliminate tri-linear terms, making the potentials similar to those for $SU(2)_L$ doublets. As we assume the VEVs of $SU(2)_L$ doublets preserve $U(1)_{em}$ the analysis of VEVs for singlets and doublets thus becomes interchangeable. For presentational simplicity we list the VEV directions in flavour space for $SU(2)_L$ singlets.

\subsection{One triplet of $\Delta(3n^2)$ or equivalently of $\Delta(6n^2)$, with $n>3$ \label{sec:6n3n1}}

The simplest potential we consider is that of one triplet of $\Delta(3n^2)$ with $n>3$, which is the same as for one triplet of $\Delta(6n^2)$ with ($n>3$). This potential wasn't studied in \cite{Ivanov:2012fp, Ivanov:2014doa}, as the renormalisable potential is invariant under a continuous symmetry (this potential has additional continuous symmetries, cf.\ Eq.~(\ref{1xD6n2_symmetries})). For $SU(2)_L$ singlets $\varphi_i$, where $i=1,2,3$ is a flavour index, the potential is
\bea
V_{\Delta(3n^2)}(\varphi) = V_{\Delta(6n^2)}(\varphi) \equiv V_0 (\varphi) 
\label{eq:potV00}
\eea 
\bea
V_0 (\varphi) =
 - ~m^2_{\varphi}\sum_i   \varphi_i \varphi^{*i}
+ r \left( \sum_i   \varphi_i \varphi^{*i}  \right)^2
 + s \sum_i ( \varphi_i \varphi^{*i})^2
 \ .
\label{eq:potV0}
\eea
For electroweak $SU(2)_L\times U(1)_Y$  doublets $H = (h_{1\alpha},h_{2\beta},h_{3\gamma})$, the respective version is
\begin{eqnarray}
V_{\Delta(3n^2)}(H) &=& V_{\Delta(6n^2)}(H) \equiv V_{0} ( H ) 
\label{eq:potV00H}
\end{eqnarray}
\begin{eqnarray}
V_{0} ( H ) &=&
 - ~m^2_{h}\sum_{i, \alpha}   h_{i \alpha}  h^{*i\alpha} + s \sum_{i, \alpha, \beta}  ( h_{i \alpha}  h^{*i\alpha})( h_{i \beta}
h^{*i\beta}) \notag \\ 
&+& \sum_{i, j, \alpha, \beta}  \left[ r_1 ( h_{i \alpha}  h^{*i\alpha})( h_{j \beta}  h^{*j \beta}) 
+ r_2 ( h_{i \alpha}  h^{*i \beta})( h_{j \beta}  h^{*j\alpha}) \right]
\  . 
\label{eq:potV0H}
\end{eqnarray}
where the greek letters denote the $SU(2)_L$ indices.

Minima are obtained analytically, and for $m_\varphi \neq 0$ the VEVs belong to four classes. As representatives we take
\begin{equation}
 (0,0,0),\ v_1\cdot(1,0,0),\ v_2\cdot(1,1,0),\ v_3\cdot(1,1,1),
 \label{1xD3n2_alignments}
\end{equation}
where 
\begin{equation}
 v_1^2=\frac{m_\varphi^2}{2r+2s},\
 v_2^2=\frac{m_\varphi^2}{4r+2s},\
 v_3^2=\frac{m_\varphi^2}{6r+2s}.
\end{equation}
There are regions of parameter space where each of these can be the global minimum.

The potential in Eq.~(\ref{eq:potV0}) can be split into two invariants with parameters $m_\varphi$ and $r$, invariant under all of $U(3)$, and the term with parameter $s$, invariant under $((U(1)\times U(1))\rtimes S_3)\times U(1)=:\Delta(6\infty^2)\times U(1)$ (the third $U(1)$ was imposed to keep the potential even and is not needed for $SU(2)_L$ doublets).
\begin{equation}
 V_0=V_{U(3)}+V_{\Delta(6\infty^2)\times U(1)}.
\end{equation}
The minima of $V_{U(3)}$ all fall into one large orbit, represented e.g.\ by $(1,0,0)$, connected to other alignments by arbitrary unitary transformations. The effect of the rest of the potential makes it less symmetric and splits the big orbit into several orbits in which the direction of the VEV becomes physical (but phases remain unphysical).

The flavour symmetries of $V_0$ (which relate each of the representative VEVs to the rest of their respective orbits) are generated by
\begin{equation}
 \begin{pmatrix}
  0&1&0\\0&0&1\\1&0&0
 \end{pmatrix},
 \begin{pmatrix}
  0&0&1\\0&1&0\\1&0&0
 \end{pmatrix},
  \begin{pmatrix}
  e^{i\alpha}&0&0\\0&e^{i\alpha}&0\\0&0&e^{i\alpha}
 \end{pmatrix}, 
  \begin{pmatrix}
  e^{i\beta}&0&0\\0&1&0\\0&0&e^{-i\beta}
 \end{pmatrix}, 
  \begin{pmatrix}
  1&0&0\\0&e^{i\gamma}&0\\0&0&e^{-i\gamma}
 \end{pmatrix},
 \label{1xD6n2_symmetries}
\end{equation}
where $\alpha,\beta,\gamma$ are arbitrary phases. Additionally, the potential is automatically invariant under canonical CP transformations, which we denotes as $CP_0$ and is associated with a unit matrix in flavour space (a $3\times3$ unit matrix in the case of one triplet) \cite{Varzielas:2016zjc}. Note that the alignments in Eq.~(\ref{1xD3n2_alignments}) all conserve canonical CP. The orbits of alignments of this potential are
\begin{equation}
 \{\begin{pmatrix}
    e^{i\eta}\\0\\0
   \end{pmatrix},\begin{pmatrix}
   0\\e^{i\eta}\\0
   \end{pmatrix},\begin{pmatrix}
    0\\0\\e^{i\eta}
   \end{pmatrix}
\},\{\begin{pmatrix}
      e^{i\eta}\\e^{i\zeta}\\0
     \end{pmatrix},\text{permut.}
\},\{\begin{pmatrix}
      e^{i\eta}\\e^{i\zeta}\\e^{i\theta}
     \end{pmatrix},\text{permut.}
\}.
\label{1xD3n2_orbits}
\end{equation}

\subsection{One triplet of $S_4$ \label{1xS4_vev_section}}

The potential of one triplet of $S_4$ is 
\begin{equation}
 V_{S_4}(\varphi)=V_0(\varphi)+V_{S_4\times U(1)}(\varphi)
\end{equation}
with
\begin{equation}
 V_{S_4\times U(1)}(\varphi)=c \left[ \left(\varphi_1 \varphi_1 \varphi^{*3} \varphi^{*3}
  +\text{cycl.}\right) + \text{h.c.} \right].
\end{equation}
\begin{eqnarray}
 V_{S_4} ( H ) =
 V_{0} (H)
+ \sum_{\alpha, \beta} \left[ c \left(
h_{1 \alpha} h_{1 \beta}  h^{*3\alpha}  h^{*3\beta} + 
\text{cycl.} \right) + \text{h.c.} \right],
\end{eqnarray}
for $SU(2)_L$ singlets and doublets respectively, and $c$ is real.

The flavour symmetries of the $S_4$ potential are generated by
\begin{equation}
 \begin{pmatrix}
  0&1&0\\0&0&1\\1&0&0
 \end{pmatrix},
  \begin{pmatrix}
  0&0&1\\0&1&0\\1&0&0
 \end{pmatrix},
  \begin{pmatrix}
  e^{i\alpha}&0&0\\0&e^{i\alpha}&0\\0&0&e^{i\alpha}
 \end{pmatrix}, 
  \begin{pmatrix}
  -1&0&0\\0&1&0\\0&0&-1
 \end{pmatrix}, 
  \begin{pmatrix}
  1&0&0\\0&-1&0\\0&0&-1
 \end{pmatrix}.
 \label{1xS4_symmetries}
\end{equation}
The potential is automatically invariant under canonical CP.
The elements of the orbits of the potential of one triplet of $\Delta(6n^2)$, Eq.~(\ref{1xD3n2_orbits}), become the following orbits:
\begin{equation}
 \begin{pmatrix}
  e^{i\eta}\\0\\0
 \end{pmatrix}\rightarrow
 \begin{pmatrix}
  1\\0\\0
 \end{pmatrix} ,
 \begin{pmatrix}
      e^{i\eta}\\e^{i\zeta}\\0
 \end{pmatrix}\rightarrow
 \begin{pmatrix}
  1\\e^{i\zeta'}\\0
 \end{pmatrix} \text{ with } \zeta'\in[0,\pi]
 \label{1xA4_orbits_from_1xD3n2_orbits_1}
\end{equation}
and
\begin{equation}
 \begin{pmatrix}
      e^{i\eta}\\e^{i\zeta}\\e^{i\theta}
 \end{pmatrix}\rightarrow
 \begin{pmatrix}
  1\\e^{i\zeta''}\\e^{i\theta'}
 \end{pmatrix} \text{ with } \zeta''\in[0,\pi] \text{ and }\theta'\in [0,2\pi].
 \label{1xA4_orbits_from_1xD3n2_orbits_2}
\end{equation}
Essentially, phases that were unphysical for $\Delta(6n^2)$ can become physical. Minimizing just the phase-dependent part, $V_{S_4\times U(1)}$, reveals the global minima
\begin{equation}
 \begin{pmatrix}
  1\\0\\0
 \end{pmatrix},
 \begin{pmatrix}
  1\\1\\1
 \end{pmatrix},
 \begin{pmatrix}
  \pm1\\\omega\\\omega^2
 \end{pmatrix},
 \begin{pmatrix}
  1\\i\\0
 \end{pmatrix},
\end{equation}
which were rephased to match their appearance in \cite{Ivanov:2014doa}.

\subsection{One triplet of $A_4$\label{1xA4_vev_section}}

The potential of one triplet of $A_4$ is an extension of the potential of one triplet of $\Delta(3n^2)$ by a term that is invariant only under $A_4\times U(1)$: 
\begin{equation}
 V_{A_4}(\varphi)=V_0(\varphi)+ V_{A_4\times U(1)}(\varphi),
\end{equation}
\begin{equation}
 V_{A_4\times U(1)}(\varphi)=c \left(
\varphi_{1} \varphi_{1} \varphi^{\ast3} \varphi^{\ast3} + \varphi_{2} \varphi_{2} \varphi^{\ast1} \varphi^{\ast1} + \varphi_{3} \varphi_{3} \varphi^{\ast2} \varphi^{\ast2}
 \right)+h.c.,
\end{equation}
\begin{eqnarray}
\label{V12H}
 V_{A_4} ( H ) =
 V_{0} (H)
+ \sum_{\alpha, \beta} \left[ c \left(
h_{1 \alpha} h_{1 \beta}  h^{*3\alpha}  h^{*3\beta} + 
\text{cycl.} \right) + \text{h.c.} \right],
\end{eqnarray}
respectively for $SU(2)_L$ singlets and doublets, and $c$ can now be complex (in contrast to $S_4$).
The full flavour-type symmetries of the full potential $V_{A_4}$ are generated by 
\begin{equation}
 \begin{pmatrix}0&1&0\\0&0&1\\1&0&0\end{pmatrix},
 \begin{pmatrix}e^{i\alpha}&0&0\\0&e^{i\alpha}&0\\0&0&e^{i\alpha}\end{pmatrix},
 \begin{pmatrix}-1&0&0\\0&1&0\\0&0&-1\end{pmatrix},
 \begin{pmatrix}1&0&0\\0&-1&0\\0&0&-1\end{pmatrix}.
 \label{1xA4_symmetries}
\end{equation}
In addition, the potential has a CP symmetry $CP_{23}$ generated e.g.\ by a CP transformation associated with the flavour space matrix \cite{Varzielas:2016zjc}
\begin{equation}
 X_{23}=\begin{pmatrix}1&0&0\\0&0&1\\0&1&0\end{pmatrix}.
 \end{equation}
Under these symmetries, the elements of the orbits of the potential of one triplet of $\Delta(6n^2)$, Eq.~(\ref{1xD3n2_orbits}), fall into the following orbits:
\begin{equation}
 \begin{pmatrix}
  e^{i\eta}\\0\\0
 \end{pmatrix}\rightarrow
 \begin{pmatrix}
  1\\0\\0
 \end{pmatrix} ,
 \begin{pmatrix}
      e^{i\eta}\\e^{i\zeta}\\0
 \end{pmatrix}\rightarrow
 \begin{pmatrix}
  1\\e^{i\zeta'}\\0
 \end{pmatrix} \text{ with } \zeta'\in[0,\pi]
 \, ,
\end{equation}
\begin{equation}
 \begin{pmatrix}
      e^{i\eta}\\e^{i\zeta}\\e^{i\theta}
 \end{pmatrix}\rightarrow
 \begin{pmatrix}
  1\\e^{i\zeta''}\\e^{i\theta'}
 \end{pmatrix} \text{ with } \zeta''\in[0,\pi] \text{ and }\theta'\in [0,2\pi].
\end{equation}

Up to two phases can become physical, which can be determined by minimizing the parts of the potential that depend on them, i.e.\  $V_{A_4\times U(1)}$.
For the alignment $(1,e^{i\zeta'},0)$,
\begin{equation}
 V_{A_4\times U(1)}[(1,e^{i\zeta'},0)]=c e^{2i\zeta'}+c^\ast e^{-2i\zeta'}, 
 \label{A4_alpha_condition}
\end{equation}
such that $\zeta'=-\text{Arg}(c)/2 \text{ mod }\pi$.
For the alignment $(  1,e^{i\zeta''},e^{i\theta'})$ we get
\begin{equation}
 (\zeta'',\theta')=(0,0),(\pi/3,5 \pi/3),(2 \pi/3,4 \pi/3),
\end{equation}
meaning we have minima $(1,1,1)$ and $(\pm 1,\omega,\omega^2)$, defining $\omega=e^{2\pi i/3}$.
Overall, we obtain the full list of possible global alignments from \cite{Ivanov:2014doa} for one triplet of $A_4$:
\begin{equation}
 \begin{pmatrix}
  1\\0\\0
 \end{pmatrix},
 \begin{pmatrix}
  1\\1\\1
 \end{pmatrix},
 \begin{pmatrix}
  \pm1\\\omega\\\omega^2
 \end{pmatrix},
 \begin{pmatrix}
  1\\e^{i\zeta'}\\0
 \end{pmatrix}.
 \label{1xA4_VEVs}
\end{equation}

\subsection{One triplet of $\Delta(27)$, or, equivalently of $\Delta(54)$ \label{sec:54271}}

The potential of one triplet of $\Delta(27)$ and $\Delta(54)$ is an extension of the $\Delta(3n^2)$ by a term that is invariant under $\Delta(54)$
\bea \label{V27P}
V_{\Delta(27)} (\varphi) &=& V_{\Delta(54)} (\varphi) =
V_0(\varphi)
+~ \left[d \left(
\varphi_1 \varphi_1 \varphi^{*2} \varphi^{*3} + 
\text{cycl.} \right) +\text{h.c.}\right]
 \ ,
\eea
and the respective $SU(2)_L$ doublet version is
\bea \label{V27H}
V_{\Delta(27)} (H) = V_{\Delta(54)} (H) = V_0(H) 
~+~ \sum_{\alpha, \beta} \left[d \left(
h_{1 \alpha} h_{1 \beta}  h^{*2 \alpha}  h^{*3 \beta} + 
\text{cycl.} \right) +
\text{h.c.}\right]
.
\eea
The potential in general violates CP explicitly. One may of course impose a CP symmetry. In \cite{Ivanov:2014doa} the two types of CP-symmetry that are normally considered consistent with the flavour-type symmetry of the potential are analysed. The 12 CP symmetries listed in \cite{Nishi:2013jqa} for $\Delta(27)$ are reduced to 6 in the context of $\Delta(54)$, e.g.\ canonical $CP_0$ and the CP symmetry associated with $X_{23}$ become related. Of the 6 remaining, 3 restrict the phase of parameter $d$ in the potential, the other 3 enforce a relation between parameters, such as $2s=(d+d^*)$ if imposing CP symmetry associated with the flavour matrix
\begin{equation}
X_4=\frac{1}{\sqrt{3}}\begin{pmatrix}1&1&1\\1&\omega&\omega^2\\1&\omega^2&\omega\end{pmatrix}
\label{X4_delta27}
\end{equation}

The full flavour-type symmetries of this potential, are generated by
\begin{equation}
 \begin{pmatrix}
  0&1&0\\0&0&1\\1&0&0
 \end{pmatrix},
  \begin{pmatrix}
  0&0&1\\0&1&0\\1&0&0
 \end{pmatrix},
  \begin{pmatrix}
  e^{i\alpha}&0&0\\0&e^{i\alpha}&0\\0&0&e^{i\alpha}
 \end{pmatrix}, 
  \begin{pmatrix}
\omega &0&0\\0&1&0\\0&0& \omega^2
 \end{pmatrix}, 
  \begin{pmatrix}
  1&0&0\\0&\omega&0\\0&0&\omega^2
 \end{pmatrix}.
 \label{1xD54_symmetries}
\end{equation}
The orbits of one triplet of $\Delta(6n^2)$, Eq.~(\ref{1xD3n2_orbits}) become 
\begin{equation}
 (1,0,0),~(1,e^{i\zeta'},0),~(1,e^{i\zeta''},e^{i\theta'}).
\end{equation}
The phases can now be physical. The phase-dependent part
\begin{equation}
 V_{\Delta(54)\times U(1)}= \left[d \left(
\varphi_1 \varphi_1 \varphi^{*2} \varphi^{*3} + 
\text{cycl.} \right) +\text{h.c.}\right]
\end{equation}
yields simply zero for the alignment $(1,e^{i\zeta'},0)$, thus $\zeta'$ remains unphysical and $(1,1,0)$ is at least a local minimum of the potential, as it was already a possible global minimum of $V_0$. For $(1,e^{i\zeta''},e^{i\theta'})$, one obtains the alignments $(1,1,1),~(1,1,\omega),(1,\omega,\omega)$ (or equivalently $(1,1,\omega^2)$). As for any value of $\text{Arg}(d)$, one of $(1,1,1),~(1,1,\omega),(1,1,\omega^2)$ makes $V_{\Delta(54)\times U(1)}$ negative, we verify that $(1,1,0)$ is never a global minimum. We have thus obtained the full list of global minima by \cite{Ivanov:2014doa}:
\begin{equation}
 (1,0,0),(1,1,1),(1,1,\omega),(1,1,\omega^2).
 \label{1xD54_VEVs}
\end{equation}

With canonical CP ($CP_0$), the last two VEVs are related whereas with the type of CP with matrix $X_4$ the last two VEVs in Eq.~(\ref{1xD54_VEVs}) become part of the same orbit and also the first two VEVs in Eq.~(\ref{1xD54_VEVs}) become part of the same orbit (separate from the last two VEVs).

\section{Potentials and some VEVs with two triplets \label{sec:two}}

\cleqn

Potentials of two triplets have two sets of terms for each triplet by themselves and also cross terms
\begin{equation}
 V(\varphi,\varphi')=V(\varphi)+V'(\varphi')+V_c(\varphi,\varphi').
\end{equation}
In the cases we consider, the two triplets transform identically under the symmetry, making $V(\varphi)$ and $V'(\varphi')$ functionally identical. 

The complete set of orbits for minima are known for the single triplet cases above, and we can proceed to two triplet potentials by analysing which degrees of freedom of $V(\varphi)+V'(\varphi')$ can become physical when the symmetry of the potential is reduced (e.g.\ by the cross-terms in $V_c(\varphi,\varphi')$). We omit the magnitudes of the VEVs, which are in general different for the two triplets.

It is convenient to define
\bea
V_1 (\varphi,\varphi') &=&
+~ \tilde r_1 \left( \sum_i \varphi_i \varphi^{*i} \right)
\left( \sum_j \varphi'_j \varphi'^{*j} \right) 
+ \tilde r_2\left( \sum_i \varphi_i \varphi'^{*i} \right)
\left( \sum_j \varphi'_j \varphi^{*j} \right) \notag \\[2mm]
&& +~ \tilde s_1\sum_i \left(\varphi_i \varphi^{*i} \varphi'_i \varphi'^{*i}
\right) \notag \\[2mm]
&& +~ \tilde s_2 \left(
\varphi_1 \varphi^{*1} \varphi'_2 \varphi'^{*2} + 
\varphi_2 \varphi^{*2} \varphi'_3 \varphi'^{*3} + 
\varphi_3 \varphi^{*3} \varphi'_1 \varphi'^{*1} 
\right)  \notag \\[2mm]
&& +~ i \, \tilde s_3 
\Big[
(\varphi_1 \varphi'^{*1} \varphi'_2 \varphi^{*2} + \text{cycl.}
) 
- 
( \varphi^{*1}\varphi'_1  \varphi'^{*2} \varphi_2 +\text{cycl.}
)
\Big],
\label{eq:potV1}
\eea
\begin{align}
 V_1(H,H')&=
\sum_{i,j,\alpha,\beta} \left[
\tilde{r}_{11}h_{i\alpha}h^{\ast i \alpha}h_{j \beta}'h'^{\ast j
  \beta}+\tilde{r}_{12}h_{i \alpha}h'^{\ast j \alpha} h'_{j \beta}h^{\ast i
  \beta}
\right] \notag \\[2mm]
 &+\sum_{i,j,\alpha,\beta} \left[
\tilde{r}_{21}h_{i\alpha}h'^{\ast i \alpha}h'_{j \beta}h^{\ast j
  \beta}+\tilde{r}_{22}h_{i \alpha}h^{\ast j \alpha}h'_{j \beta}h'^{\ast i
  \beta}
\right]\notag \\[2mm]
 &+\sum_{i,\alpha,\beta} \left[
\tilde{s}_{11}h_{i \alpha}h^{\ast i \alpha}h'_{i \beta}h'^{\ast i
  \beta}+\tilde{s}_{12}h_{i \alpha}h'^{\ast i \alpha}h'_{i \beta}h^{\ast i
  \beta}
\right]\notag \\[2mm]
 &+\sum_{\alpha,\beta} \left[
\tilde{s}_{21}(h_{1\alpha}h^{\ast1\alpha}h'_{2\beta}h'^{\ast2\beta}+\text{cycl.})+\tilde{s}_{22}(h_{1\alpha}h'^{\ast
  2 \alpha}h'_{2\beta}h^{\ast 1\beta}+\text{cycl.})
\right]\notag \\[2mm]
 &+i\tilde{s}_{31}\sum_{\alpha,\beta} 
[(h_{1\alpha}h'^{\ast1\alpha}h'_{2\beta}h^{\ast2\beta}+\text{cycl.}) - (h^{\ast1\alpha}h'_{1\alpha}h'^{\ast2\beta}h_{2\beta}+\text{cycl.})]\notag \\[2mm]
 &+i\tilde{s}_{32}\sum_{\alpha,\beta} [(h_{1\alpha}h^{\ast 2 \alpha}h'_{2\beta}h'^{\ast1\beta}+\text{cycl.}) - (h^{\ast1\alpha}h_{2 \alpha}h'^{\ast2\beta}h'_{1\beta}+\text{cycl.})],
\label{eq:potV1H}
 \end{align}
\bea\label{eq:potV2}
V_2 (\varphi,\varphi') &=&
 \tilde r_1 \left( \sum_i \varphi_i \varphi^{*i} \right)
\left( \sum_j \varphi'_j \varphi'^{*j} \right) 
+ \tilde r_2\left( \sum_i \varphi_i \varphi'^{*i} \right)
\left( \sum_j \varphi'_j \varphi^{*j} \right) \notag \\[2mm]
&&
 +~ \tilde s_1\sum_i \left(\varphi_i \varphi^{*i} \varphi'_i \varphi'^{*i}
\right), 
\eea
\bea
 V_2(H,H')&=&\sum_{i,j,\alpha,\beta}\left[
\tilde{r}_{11}h_{i\alpha}h^{\ast i \alpha}h_{j \beta}'h'^{\ast j
  \beta}+\tilde{r}_{12}h_{i \alpha}h'^{\ast j \alpha} h'_{j \beta}h^{\ast i
  \beta} \right]\notag \\
& &+\sum_{i,j,\alpha,\beta}\left[
\tilde{r}_{21}h_{i\alpha}h'^{\ast i \alpha}h'_{j \beta}h^{\ast j \beta}+\tilde{r}_{22}h_{i \alpha}h^{\ast j \alpha}h'_{j \beta}h'^{\ast i \beta}\right]\notag \\
 &&+\sum_{i,\alpha,\beta}\left[
\tilde{s}_{11}h_{i \alpha}h^{\ast i \alpha}h'_{i \beta}h'^{\ast i \beta}+\tilde{s}_{12}h_{i \alpha}h'^{\ast i \alpha}h'_{i \beta}h^{\ast i \beta}\right].
\label{eq:potV2H}
 \eea

\subsection{Two triplets of $\Delta(6n^2)$ with $n>3$ \label{sec:6n22}}

The potentials of two triplets of $\Delta(6n^2)$ are, using the definitions of $V_0$ and $V_2$ previously,
\bea
V_{\Delta(6n^2)} (\varphi,\varphi')&=& V_0(\varphi)  + V'_0(\varphi')  + V_2(\varphi,\varphi')  \ , \label{V6n2PP} \\[2mm]
V_{\Delta(6n^2)} (H,H')&=& V_0(H) + V'_0(H') + V_2(H,H')\ \label{V6n2HH} .
\eea
The orbits for one triplet of $\Delta(6n^2)$ are in Eq.~(\ref{1xD3n2_orbits}),
and we can obtain minima for the two triplet case by combining any two members of any orbit (not just the representatives), and then check which phases are unphysical.
The symmetries of the two triplet potential are generated by simultaneous transformations of both triplets under $\Delta(6\infty^2)$ and by separate $U(1)$ phases acting on each triplet,
\begin{equation}
\begin{pmatrix}
e^{i\alpha}&&\\&e^{i\alpha}&\\&&e^{i\alpha} 
\end{pmatrix}\oplus
\begin{pmatrix}
1&&\\&1&\\&&1 
\end{pmatrix}\text{, and }
\begin{pmatrix}
1&&\\&1&\\&&1 
\end{pmatrix}\oplus
\begin{pmatrix}
e^{i\alpha'}&&\\&e^{i\alpha'}&\\&&e^{i\alpha'} 
\label{two_triplet_phases}
\end{pmatrix},
\end{equation}
as well as an overall canonical CP transformation. Accounting for these, we get the following combinations of orbits:
\begin{align}
 (e^{i\eta},0,0),(e^{i\eta'},0,0)&\rightarrow (1,0,0),(1,0,0)\label{2xDelta6n2vevsstart}\\
 (e^{i\eta},0,0),(0,e^{i\eta'},0)&\rightarrow (1,0,0),(0,1,0)\\
 (e^{i\eta},0,0),(e^{i\eta'},e^{i\zeta'},0)&\rightarrow (1,0,0),(1,1,0)\\
 (e^{i\eta},0,0),(0,e^{i\eta'},e^{i\zeta'})&\rightarrow (1,0,0),(0,1,1)\\
 (e^{i\eta},0,0),(e^{i\eta'},e^{i\zeta'},e^{i\theta'})&\rightarrow (1,0,0),(1,1,1)\\
  (e^{i\eta},e^{i\zeta},0),(0,e^{i\zeta'},e^{i\theta'})&\rightarrow (1,1,0),(0,1,1)\label{2xDelta6n2vevsend}\\
 (e^{i\eta},e^{i\zeta},0),(e^{i\eta'},e^{i\zeta'},0)&\rightarrow (1,1,0),(1,e^{i\zeta'},0)\\
 (e^{i\eta},e^{i\zeta},0),(e^{i\eta'},e^{i\zeta'},e^{i\theta'})&\rightarrow (1,1,0),(1,e^{i\zeta'},1)\\
 (e^{i\eta},e^{i\zeta},e^{i\theta}),(e^{i\eta'},e^{i\zeta'},e^{i\theta'})&\rightarrow (1,1,1),(1,e^{i\zeta'},e^{i\theta'}).
\end{align}
The phases can now be fixed by the minimisation of the phase-dependent part of the potential. We get for $(1,1,0),(1,e^{i\zeta'},0)$ and for $(1,1,0),(1,e^{i\zeta'},1)$ that $\zeta'=0$ for $r_2'>0$ and $\zeta'=\pi$ for $r_2<0$, i.e.:
\begin{equation}
 (1,1,0),(1,\pm1,0) \text{ and } (1,1,0),(1,\pm1,1)
\end{equation}
different sign choices are different orbits.
For $(1,1,1),(1,e^{i\zeta'},e^{i\theta'})$, we get for $r_2<1$ the orbit
\begin{equation}
 (1,1,1),(1,1,1)\label{2xD6n2_111_111}
\end{equation}
and for $r_2>0$ the orbits
\begin{equation}
 (1,1,1),(1,\omega,\omega^2)\text{ and }(1,1,1),(1,\omega^2,\omega).\label{2xD6n2_111_1omom}
\end{equation}

\subsection{Two triplets of $\Delta(3n^2)$ with $n>3$ \label{sec:3n22}}

The potentials for two triplets of $\Delta(3n^2)$ with $n>3$ are
\bea
V_{\Delta(3n^2)} (\varphi,\varphi') &=&
V_0 (\varphi) + V_0'(\varphi') + V_1 (\varphi, \varphi'),
\label{V3n2PP}
\eea
\bea
 V_{\Delta(3n^2)} (H,H') &= V_0(H) + V_0'(H')+ V_1(H,H')\ .\label{V3n2HH}
\eea
The single triplet orbits of $\Delta(3n^2)$ and $\Delta(6n^2)$ were the same and listed in Eq.~(\ref{1xD3n2_orbits}).
The difference to the previous potential lies in the fact that the full symmetries of $V_{\Delta(3n^2)}$ only allow for cyclic permutations, i.e.\ only
\begin{equation}
 \begin{pmatrix}
  &1&\\&&1\\1&&
 \end{pmatrix}\oplus
 \begin{pmatrix}
  &1&\\&&1\\1&&
 \end{pmatrix},
\end{equation}
in addition to all phase symmetries arising arising from $\Delta(3n^2)$ and Eq.~(\ref{two_triplet_phases}).
This two triplet potential has no automatic CP symmetry.
Compared to $\Delta(6n^2)$, several orbits split,
but interchanging the first and second triplet allow us to reduce the number of distinct such orbits to $(1,0,0),(1,1,0)$ and $(1,0,0),(1,0,1)$.
 Besides Eqs.\ (\ref{2xDelta6n2vevsstart})--(\ref{2xDelta6n2vevsend}), Eqs.\ (\ref{2xD6n2_111_111}),(\ref{2xD6n2_111_1omom}), and the two orbits above that arise from splitting known orbits, two other new orbits can arise due to the lacking CP symmetry, namely
\begin{equation}
 (1,1,0),(1,e^{i\zeta'},0) \text{ and } (1,1,0),(1,e^{i\zeta'},1)
 \end{equation}
with $\zeta'=\text{arctan}(\tilde r_2/\tilde s_3)$ a function of $\tilde s_3$ and $\tilde r_2$ in contrast to the situation with a $\Delta(6n^2)$ symmetry, where $\zeta'=0,\pi$, depending on the value of $\tilde r_2$.
The pair $(1,1,0),(1,e^{i\zeta'},0)$ and the pair $(1,1,0),(1,e^{i\zeta'},1)$ have the same $\zeta'$. When special CP symmetries are imposed, then $\zeta'$ can be forced to take special values again.

\subsection{Two triplets of $S_4$ \label{sec:S4PPVEVs}}

The potentials for two triplets of $S_4$ are
\bea
V_{S_4} (\varphi,\varphi') &=&
V_0 (\varphi) + V'_0 (\varphi') + V_2 (\varphi,\varphi') +  \label{V24PP}\\[2mm]
&&+~
c \left[\left(
\varphi_1 \varphi_1 \varphi^{*3} \varphi^{*3} + 
\text{cycl.}\right)+\text{h.c.}\right]
 + c' \left[\left(
\varphi'_1 \varphi'_1 \varphi'^{*3} \varphi'^{*3} + 
\text{cycl.} \right)+ \text{h.c.}\right]\notag\\[2mm]
& &+ ~\tilde c \left[ \left(
\varphi_1 \varphi'_1 \varphi^{*3} \varphi'^{*3} + 
\text{cycl.}\right) + \text{h.c.}\right], \notag
\eea
\begin{align}
V_{S_4} (H,H') =~& V_0(H) + V_0'(H')+ V_2(H,H') \label{V24HH}\\
&+\sum_{\alpha,\beta} c\left[  \left(
h_{1\alpha} h_{1\beta} h^{\ast3\alpha} h^{\ast3\beta} + 
\text{cycl.} \right)+ \text{h.c.}\right]
+\sum_{\alpha,\beta}  c' \left[\left(
h'_{1 \alpha} h'_{1 \beta} h'^{\ast3\alpha} h'^{\ast3\beta} + 
\text{cycl.}\right) + \text{h.c.} \right]
\notag\\
&+\sum_{\alpha,\beta} \tilde{c}_1\left[ \left(h_{1\alpha}h^{\ast3
    \alpha}h'_{1\beta}h'^{\ast3\beta}+ \text{cycl.}\right)+\text{h.c.}\right]
+\sum_{\alpha,\beta} \tilde{c}_2\left[\left(h_{1\alpha}h'^{\ast 3\alpha}h'_{1\beta}h^{\ast3\beta}+\text{cycl.}\right)+\text{h.c.} \right] . \notag
\end{align}
For one triplet, the symmetry generators are in Eq.~(\ref{1xS4_symmetries}),
and the potential has an automatic CP symmetry. The single triplet orbits are
\begin{equation}
 \{\begin{pmatrix}\pm e^{i\alpha}\\0\\0\end{pmatrix}\},
 \{\begin{pmatrix}(-1)^k e^{i\alpha}\\(-1)^l e^{i\alpha}\\(-1)^{k+l} e^{i\alpha}\end{pmatrix}\},
 \{\begin{pmatrix}(-1)^k e^{i\alpha}\\\omega(-1)^l e^{i\alpha}\\\omega^2(-1)^{k+l} e^{i\alpha}\end{pmatrix}\},
 \{\begin{pmatrix}-(-1)^k e^{i\alpha}\\\omega(-1)^l e^{i\alpha}\\\omega^2(-1)^{k+l} e^{i\alpha}\end{pmatrix}\},
 \{\begin{pmatrix}0\\\pm e^{i \alpha}\\\pm i e^{i\alpha}\end{pmatrix}\},
\end{equation}
where the sign choices represent independent orbits.
We combine the single triplet orbits to obtain the two triplet orbit representatives
\begin{align}
(1,0,0),(1,0,0)\\
(1,0,0),(0,1,0)\\
(1,0,0),(1,0,i)\\
(1,0,0),(0,1,i)\\
(1,0,0),(1,1,1)\\
(1,0,0),(1,\omega^2,\omega)\\
(1,0,i),(1,0,\pm i)\\
(1,0,i),(1,i,0)\\
(1,0,i),(1,1,1)\\
(1,0,i),(1,\omega^2,\pm\omega)\\
(1,1,1),(1,1,\pm1)\\
(1,1,1),(1,\pm\omega^2,\omega)\\
(1,\omega^2,\omega),(1,\omega^2,\pm\omega)\\
(1,\omega^2,\omega),(1,-\omega,-\omega^2)\\
(1,\omega^2,\omega),(1,\omega,\omega^2)
\end{align}

\subsection{Two triplets of $A_4$ \label{sec:A4PPVEVs}}

The potentials for two triplets of $A_4$ are
\bea
\label{V12PP}
V_{A_4} (\varphi,\varphi') &=&
V_0 (\varphi) + V'_0 (\varphi') + V_1 (\varphi,\varphi') +  \\[2mm]
&&+ \left[
c \left(
\varphi_1 \varphi_1 \varphi^{*3} \varphi^{*3} + 
\text{cycl.} \right) +\text{h.c.}\right]
 + \left[ c' \left(
\varphi'_1 \varphi'_1 \varphi'^{*3} \varphi'^{*3} + 
\text{cycl.} \right)+\text{h.c.}\right]\notag\\[2mm]
 &&+ \left[\tilde c \left(
\varphi_1 \varphi'_1 \varphi^{*3} \varphi'^{*3} + 
\text{cycl.} \right) + \text{h.c.}
\right] ,\notag
\eea
\begin{align}
\label{V12HH}
V_{A_4} (H,H') &= V_0(H) + V_0'(H')+ V_1(H,H')\\
&+\sum_{\alpha,\beta}\left[ c \left(
h_{1\alpha} h_{1\beta} h^{\ast3\alpha} h^{\ast3\beta} + 
\text{cycl.} \right)
+ c' \left(
h'_{1 \alpha} h'_{1 \beta} h'^{\ast3\alpha} h'^{\ast3\beta} + 
\text{cycl.} \right) + \text{h.c.} \right]
\notag\\[2mm]
&+\sum_{\alpha,\beta}\left[ \tilde{c}_1(h_{1\alpha}h^{\ast3 \alpha}h'_{1\beta}h'^{\ast3\beta}+ \text{cycl.})+\tilde{c}_2(h_{1\alpha}h'^{\ast 3\alpha}h'_{1\beta}h^{\ast3\beta}+\text{cycl.})+\text{h.c.} \right] \notag.
\end{align}
The symmetry generators for one triplet are in Eq.~(\ref{1xA4_symmetries}), and the orbit representatives are
\begin{equation}
 \{\begin{pmatrix}\pm e^{i\alpha}\\0\\0\end{pmatrix}\},
 \{\begin{pmatrix}(-1)^k e^{i\alpha}\\(-1)^l e^{i\alpha}\\(-1)^{k+l} e^{i\alpha}\end{pmatrix}\},
 \{\begin{pmatrix}(-1)^k e^{i\alpha}\\\omega(-1)^l e^{i\alpha}\\\omega^2(-1)^{k+l} e^{i\alpha}\end{pmatrix}\},
 \{\begin{pmatrix}-(-1)^k e^{i\alpha}\\\omega(-1)^l e^{i\alpha}\\\omega^2(-1)^{k+l} e^{i\alpha}\end{pmatrix}\},
 \{\begin{pmatrix}0\\\pm e^{i \alpha}\\\pm e^{i\alpha+i\beta}\end{pmatrix}\},
\end{equation}
For two triplets we get
\begin{align}
 (1,0,0),(1,0,0)\\
(1,0,0),(0,1,0)\\
(1,0,0),(1,e^{i \alpha'},0)\\
(1,0,0),(0,1,e^{i \alpha'})\\
(1,0,0),(e^{i \alpha'},0,1)\\
(1,0,0),(1,1,1)\\
(1,0,0),(1,\omega,\omega^2)\\
(1,e^{i \alpha},0),(1,\pm e^{i \alpha'},0)\\
(1,e^{i \alpha},0),(0,1,e^{i \alpha'})\\
(1,e^{i \alpha},0),(e^{i \alpha'},0,1)\\
(1,e^{i \alpha},0),(1,\pm1,1)\\
(1,e^{i \alpha},0),(1,\pm\omega,\omega^2)\\
(1,1,1),(1,1,\pm1)\\
(1,1,1),(1,\omega,\pm\omega^2)\\
(1,1,1),(1,\omega,\omega^2)\\
(1,\omega,\omega^2),(1,\omega,\pm\omega^2)\\
(1,\omega,\omega^2),(1,\omega,\omega^2)
\end{align}
where $\alpha$ and $\alpha'$ are fixed by the respective one-triplet parts of the two-triplet potential, as in Eq.~(\ref{A4_alpha_condition}). Note that
$(1,0,0),(e^{i \alpha'},1,0)$ as well as 
$(1,0,0),(0,e^{i \alpha'},1)$ and
$(1,0,0),(1,0,e^{i \alpha'})$,
are part of the above orbits due to the separate rephasing symmetries of each triplet.

\subsection{Two triplets of $\Delta(54)$ \label{sec:D54PPVEVs}}

The potentials for two triplets of $\Delta(54)$ are
\begin{align}
 V_{\Delta(54)} (\varphi,\varphi')=&V_0(\varphi)+V'_0(\varphi')+V_2(\varphi,\varphi') \label{V54PP}\\&+\left[d \left(
\varphi_1 \varphi_1 \varphi^{*2} \varphi^{*3} + 
\text{cycl.} \right) + \text{h.c.}\right] + \left[d' \left(
\varphi'_1 \varphi'_1 \varphi'^{*2} \varphi'^{*3} + 
\text{cycl.} \right) +\text{h.c.}\right]
 \notag\\&+\tilde d_1\left[   \left(
\varphi_1 \varphi'_1 \varphi^{*2} \varphi'^{*3} + 
\text{cycl.} \right) +\left(
\varphi_1 \varphi'_1 \varphi^{*3} \varphi'^{*2} + 
\text{cycl.} \right) \right]+\text{h.c.}, \notag
\end{align}
\begin{align} 
 V_{\Delta(54)} (H,H') &=V_0(H) + V_0'(H')+ V_2(H,H') \label{V54HH}\\&+ 
  \sum_{\alpha,\beta}\left[ d \left(
h_{1\alpha} h_{1\beta} h^{\ast 2 \alpha} h^{\ast 3 \beta}+
\text{cycl.} \right) + 
d' \left(
h'_{1\alpha} h'_{1\beta} h'^{\ast 2 \alpha} h'^{\ast 3 \beta} + 
\text{cycl.} \right)
+ \text{h.c.}\right] \notag
 \\&+  \sum_{\alpha,\beta} \left[
 \tilde{d}_{11}(h_{1\alpha}h^{\ast 2 \alpha}h'_{1\beta}h'^{\ast 3 \beta} +
\text{cycl.})+\tilde{d}_{12}(h_{1\alpha}h'^{\ast 3 \alpha}h'_{1\beta}h^{\ast 2 \beta} +
\text{cycl.}) + \text{h.c.} \right]  \notag
 \\&+  \sum_{\alpha,\beta} \left[  \tilde{d}_{11}(h_{1\alpha}h^{\ast 3 \alpha}h'_{1\beta}h'^{\ast 2 \beta}+
\text{cycl.})+\tilde{d}_{12}(h_{1\alpha}h'^{\ast 2 \alpha}h'_{1\beta}h^{\ast 3 \beta}+
\text{cycl.}) + \text{h.c.} \right] \! . ~~ \notag
\end{align}
This potential has no automatic CP symmetries.
We can write
\begin{align}
 V_{\Delta(54)}(\varphi,\varphi')=&V_{\Delta(6n^2)}(\varphi)+V'_{\Delta(6n^2)}(\varphi')+V_{\Delta(54)}(\varphi)+V'_{\Delta(54)}(\varphi')\\&+V_{c,\Delta(6n^2)}(\varphi,\varphi')+V_{c,\Delta(54)}(\varphi,\varphi').
\end{align}
The orbits of VEVs for one triplet are
\begin{equation}
 \{\begin{pmatrix}\omega^k e^{i\alpha}\\0\\0\end{pmatrix},\text{perm.}\},\{\begin{pmatrix}\omega^k e^{i\alpha}\\\omega^l e^{i\alpha}\\\omega^{2k+2l} e^{i\alpha}\end{pmatrix},\text{perm.}\},\{\begin{pmatrix}\omega^k e^{i\alpha}\\\omega^l e^{i\alpha}\\\omega^{2k+2l+1} e^{i\alpha}\end{pmatrix},\text{perm.}\},\{\begin{pmatrix}\omega^k e^{i\alpha}\\\omega^l e^{i\alpha}\\\omega^{2k+2l+2} e^{i\alpha}\end{pmatrix},\text{perm.}\}.
 \label{delta54_orbits}
\end{equation}
In addition to the direct sum of the generators in Eq.~(\ref{1xD54_symmetries}), the potential is invariant under separate phase symmetries for each triplet, as in Eq.~(\ref{two_triplet_phases}). By combining single triplet orbits and then eliminating unphysical degrees of freedom we get
\begin{align}
 &(1,0,0),(1,0,0) \label{2XD54_1}\\
 &(1,0,0),(0,1,0)\\ 
 &(1,0,0),(1,1,1)\\ 
 &(1,0,0),(1,1,\omega)\\ 
 &(1,0,0),(1,1,\omega^2)\\
 &(1,1,\omega^i),(\omega^{k'-k},\omega^{l'-l},\omega^{2k'+2l'-2k-2l+i'}) 
 \label{2XD54_vevs}
 \end{align}
where the last case has several orbits labeled by $i$ and $i'$. We note that phase differences between the two triplets are physical.

\subsection{Two triplets of $\Delta(27)$ \label{sec:D27PPVEVs}}

The potentials for two triplets of $\Delta(27)$ are 
\bea\label{V27PP}
V_{\Delta(27)} (\varphi,\varphi') &\!\!=\!\!&
V_0 (\varphi) + V_0'(\varphi') + V_1 (\varphi, \varphi') 
  \\[2mm]
&& + \left[d \left(
\varphi_1 \varphi_1 \varphi^{*2} \varphi^{*3} + 
\text{cycl.} \right) + \text{h.c.}\right] + \left[d' \left(
\varphi'_1 \varphi'_1 \varphi'^{*2} \varphi'^{*3} + 
\text{cycl.} \right) +\text{h.c.}\right]
\notag\\[2mm]
&& +\left[ \tilde d_1  \left(
\varphi_1 \varphi'_1 \varphi^{*2} \varphi'^{*3} + 
\text{cycl.} \right) +\text{h.c.}\right] 
+ \left[ \tilde d_2  \left(
\varphi_1 \varphi'_1 \varphi^{*3} \varphi'^{*2} + 
\text{cycl.} \right) +\text{h.c.}\right]
\notag 
 \!,
\eea
\bea
 V_{\Delta(27)} (H,H') \!\!&\!\!\!\!=\!\!\!\!& \!\!V_0(H) + V_0'(H')+ V_1(H,H')+ \label{V27HH}\\
& +&\!\!\sum_{\alpha,\beta}\left[ d \left(
h_{1\alpha} h_{1\beta} h^{\ast 2 \alpha} h^{\ast 3 \beta}+
\text{cycl.} \right) + 
d' \left(
h'_{1\alpha} h'_{1\beta} h'^{\ast 2 \alpha} h'^{\ast 3 \beta} + 
\text{cycl.} \right)
+ \text{h.c.}\right]
\notag\\
&+& \!\!\sum_{\alpha,\beta} \left[
 \tilde{d}_{11}(h_{1\alpha}h^{\ast 2 \alpha}h'_{1\beta}h'^{\ast 3 \beta} +
\text{cycl.})+\tilde{d}_{12}(h_{1\alpha}h'^{\ast 3 \alpha}h'_{1\beta}h^{\ast 2 \beta} +
\text{cycl.}) + \text{h.c.} \right] \notag \\
&+&\!\! \sum_{\alpha,\beta} \left[  \tilde{d}_{21}(h_{1\alpha}h^{\ast 3 \alpha}h'_{1\beta}h'^{\ast 2 \beta}+
\text{cycl.})+\tilde{d}_{22}(h_{1\alpha}h'^{\ast 2 \alpha}h'_{1\beta}h^{\ast 3 \beta}+
\text{cycl.}) + \text{h.c.} \right] \notag
 \! .
\eea
The one triplet VEVs for $\Delta(27)$ are the same as for $\Delta(54)$, so the VEV pairs generated by our method are similar to the case of two triplets of $\Delta(54)$. There is however the permutation generator that is missing in $\Delta(27)$, and therefore several orbits split with respect to $\Delta(54)$. In addition to Eqs.~(\ref{2XD54_1}-\ref{2XD54_vevs}), we get independent orbits:
\begin{align}
 &(1,0,0),(1,1,\omega)\\ 
 &(1,0,0),(1,\omega,1)\\ 
 &(1,0,0),(1,1,\omega^2)\\
 &(1,0,0),(1,\omega^2,1).
 \end{align}

\section{Conclusions \label{sec:con}}

In this paper we have analysed the minima of scalar potentials for multi-Higgs models, 
where the scalars are arranged as either one triplet or two triplets of the discrete symmetries
$\Delta(3n^2)$ and $\Delta(6n^2)$ with $n=2$ ($A_4$, $S_4$), $n=3$ ($\Delta(27)$, $\Delta(54)$) and $n>3$.
We have found the minima with a technique where we consider by steps the symmetry of parts of the potential and progressively add terms that reduce the symmetry, minimizing them in turn. Whether the minima spontaneously violate CP or not will be discussed in a future work \cite{future}.
The results should be useful for both multi-Higgs models involving 
electroweak doublets and multi-flavon models involving electroweak singlets,
where in both cases the fields transform as triplets under some non-Abelian discrete symmetry.

\section*{Acknowledgements}
IdMV acknowledges funding from Funda\c{c}\~{a}o para a Ci\^{e}ncia e a Tecnologia (FCT) through the
contract IF/00816/2015.
This project has received funding from the European Union's Seventh Framework Programme for research, technological development and demonstration under grant agreement no PIEF-GA-2012-327195 SIFT.

The work of CL is
supported by the Deutsche Forschungsgemeinschaft (DFG) within the
Research Unit FOR 1873 ``Quark Flavour Physics and Effective Field
Theories''.

SFK acknowledges the STFC Consolidated Grant ST/L000296/1 and the European Union's Horizon 2020 Research and Innovation programme under Marie Sk\l{}odowska-Curie grant agreements Elusives ITN No.\ 674896 and InvisiblesPlus RISE No.\ 690575. 

Work  supported  by  MINECO  grants  FPA2014-58183-P, Multidark CSD2009-00064, and the PROMETEOII/2014/084 grant from Generalitat Valenciana.

\end{document}